# A Sustainable approach to large ICT Science based infrastructures; the case for Radio Astronomy

Domingos Barbosa [#1], João Paulo Barraca [#&], Albert-Jan Boonstra [*], Rui Aguiar [#&], Arnold van Ardenne [*£], Juande de Santander-Vela [§], Lourdes Verdes-Montenegro [§]

[#] *Instituto de Telecomunicações, Campus Universitário de Santiago,*

*3810-193 Aveiro, Portugal*
[1] `dbarbosa@av.it`

[&] *Universidade de Aveiro, Campus Universitário de Santiago,*

*3810-193 Aveiro, Portugal*

[*] *ASTRON, P.O. Box 2, 7990 AA Dwingeloo,*
*The Netherlands*

[£] *Chalmers University of Technology, Gothenburg,*
*Sweden*

[§] *Instituto de Astrofísica de Andalucía (IAA-CSIC)*
*Glorieta de la Astronomía s/n, E-18008,*
*Granada, Spain*

*Abstract*—Large sensor-based infrastructures for radio astronomy will be among the most intensive data-driven projects in the world, facing very high power demands. The geographically wide distribution of these infrastructures and their associated processing High Performance Computing (HPC) facilities require Green Information and Communications Technologies (ICT): a combination is needed of low power computing, power and byte efficient data storage, local data services, Smart Grid power management, and inclusion of Renewable Energies. Here we outline the major characteristics and innovation approaches to address power efficiency and long-term power sustainability for radio astronomy projects, focusing on Green ICT for science.

I. INTRODUCTION

The Energy Sustainability of large-scale scientific infrastructures led to consider the impact of their carbon footprint and Power costs into the respective development path and lifetimes [1]. Additionally, the Roadmap of the European Strategy Forum on Research Infrastructures (ESFRI) [2] has indicated that it is paramount that a multitude of test facilities and Research Infrastructures should lead the world in the efficient use of energy, promote new renewable forms of energy, and develop low carbon emission technologies, to be adopted as part of a future Strategic Energy Technology Plan. Radio astronomy projects will be among the most data-intense and power hungry projects. Recent experiences with Square Kilometer Array (SKA) [1] precursors and pathfinders like ASKAP, MeerKAT and LOFAR reveal that an important part of the life cycle cost of these large-scale radio astronomy projects will be power consumption [6],[7]. As an example, a 30-meter radio telescope requires approximately 50 kW during operation (about 1GWh for a typical 6h VLBI - observation experiment) enough to power a small village, while new infrastructures based on Aperture Arrays, promising huge sky survey speeds, may require even more, based on estimated digital processing needs [16]. By many considered an ICT based infrastructure due to its emphasis on intense data processing and high performance computing at exa-scale level, the SKA design emphasis is driving compute power while requiring to closely watch power consumption for societal and operational cost reasons (OPEX). When completed, SKA will therefore set the highest constrains on power consumption and availability among all science infrastructures, surpassing current figure considerably as can be perceived by Table I.

Addressing both the reduction of electricity costs and the generation and management of electricity is paramount to avoid future inefficiencies and higher costs. For instance, the Atacama Large Millimeter Array (ALMA) interferometer and the Very Large Telescope (VLT) in the Chilean Andes are powered from diesel generators, leading now the European Southern Observatory (ESO) to consider greener energy sources to its Very Large Telescope (VLT) facilities in Paranal [8]. At this site, electrical power is produced in off-grid mode using a combination of efficient multi-fuel turbine generators (2.6MWe at the site) that can use fuel sources with lower carbon footprint like natural gas, or Liquefied Petroleum Gas (LPG), combined with diesel generators connected to a 10kV power grid. However, electricity prices in Chile rose on average by 7% per year between 2003 and 2010 according to statistics from the Organization for Economic Cooperation and Development (OECD) [8],[14]. Therefore, the ALMA permanent power system plant, capable of providing up to 7MW peak in "island -mode" is already prepared to connect to a renewable power plant, and the European Extremely Large Telescope (E-ELT) might include options for renewables when market options in Chile make these technologies economically accessible [17]. Hence, fossil fuel price fluctuations and longer term availability and associated price rises represent a challenge in terms of planning a suitable energy mix supply, in particular for remotely located infrastructures. SKA itself developed Power Investigation strategies in order to identify avenues for new power efficiencies and sustainable power provision. Naturally, this reflection has prompted the need to address the new trends in green ICT and the necessity to evaluate carefully through demo projects new smart grid and renewable options, in particular solar power. These facts have prompted ASTRON in the context of the Peta byte compute level Low Frequency Array (LOFAR) now in operation, to embark on a low-power intense computing and data processing evaluation, and to develop a solar energy supply research strategy in a wider European collaboration. This collaboration under the framework of the novel SKA Aperture Array technologies, will investigate for instance renewables inclusion with concentrated solar thermal, in off-grid mode and storage capacity for night operation, through real-scale testing in Portugal, among several power options. Specifically, the BIOSTIRLING-4SKA project [3],[14]-[16] started in 2013 to study the cost viability, reliability and life-time of solar thermal concentrator pilot plant with Stirling Dish engines for a100kW production with bioenergy storage for continuous operation of radio astronomy infrastructures. The real-scale technology prototypes will be deployed in southeast Portugal, where earlier SKA-related prototypes have already been tested. The site combines one of the lowest Radio Frequency Interference levels in Europe, with one the most solar-intense geographical areas of Europe. These characteristics make it ideal to optimally test radio astronomy prototypes powered by solar energy [19].

TABLE I
SKA POWER PROJECTIONS FOR DIFFERENT SUBSYSTEMS AND PHASES; TARGET: **< 100MW**

| SKA Phase 1&2 | South Africa | Australia |
|---|---|---|
| Sparse Arrays | | 3.36 MW |
| Mid Dishes | 2.5MW | |
| Survey Dishes | | 1.2MW |
| On-site Computing | 4.7MW | 1.32MW |
| Totals/site | 5.7MW | 4.8MW |
| SKA Phase2 incl. Dense Arrays | ~80MW (SKA Phase 2 configuration not known yet) | |
| Off-site Computing | ~30-40MW (SKA Phase 2 configuration not known yet) | |

II. GREEN ICT : FROM CORRELATORS TO CLOUDS

Green ICT can defined as embodying "design, manufacturing, utilization, disposal of computers, servers, and associated subsystems—such as monitors, printers, massive storage devices, and networking and communications systems — efficiently and effectively and with minimal or no impact on the environment" [31]. Hence, Green ICT paved the way towards more efficient intense computing systems. In fact, the biggest computing challenge within radio astronomy lies within the architecture of the correlator of big synthesis radio telescopes and the second tier processing and storage infrastructures. The correlator processes the data streams arising from the large number of antenna elements of say, with N>1000 antennas. The optimum architecture is planned to minimize power consumption as much as possible by following several approaches: minimizing I/O (storage media, and network interconnects) and memory operations, implying preference for a matrix structure over a pipeline structure and avoiding the use of memory banks and choose among the lowest power computing chip technology. For instance, the ALMA correlator selected for its core design the StratixII 90nm technology based on considerations on power dissipation and logic resources while much lower power technologies are available now. The SKA, under the Central Signal Processor Element Consortium, is currently developing design concepts for design for N>2000 and over 1 GHz frequency bandwidth, based on Application-specific integrated circuits (ASICs) fabricated in a 20nm CMOS process, still better than 20nm for FPGAS with low power considerations. Excluding antenna data pre-processing, the SKA correlator is estimated to consume less than 100 kW [24].

After data is integrated by the correlator and further processed to create calibrated data, it must be stored in a permanent media, such as the case of massive Storage Area Networks (SANs), relying in rotational technologies such as hard disks. ALMA can output several TeraBytes of data per project that must be stored, and the future SKA infrastructure is expected to produce close to an Exabyte/day of raw information, prior to further processing and data reduction. All these data must be made available in large facilities for further reduction by researchers (eg, using CASA [26]). Due to the amount of information, and the costs of transmitting data through long distance optical links, it is paramount the use of computation facilities located in close proximity to the source of information, but also close to researchers, in order to reduce latency and cost of the post-analysis process.

The typical approach is to create computational behemoths capable of handling the entire operation of the instruments, storage, and frequently further processing of the data produced. However, especially during the first years of operation, large infrastructures are operating with frequent interruptions caused by detection of erroneous or unexpected behavior, or when operations require further tuning. Even after entering into its normal operational status, instruments are, among other factors, affected by maintenance downtime, and also by weather conditions limiting observations. As an example, according to the ALMA cycle 0 report, over the course of 9 months (total of ~6500 hours), the instrument was allocated for 2724 hours of observation time, and this resulted in 38% (1034 hours) of successful observation [22],[24]. This results in a considerable efficiency loss, considering all the processing infrastructure that must be available, independent of the observation status. Although we believe the initial processing must be done close to the location of the sensors, we also believe that processing should be shared or co-located as much as possible to other already existing infrastructures, exploiting time multiplexing as a way of increasing power efficiency. Moreover, further offline reduction methods can be improved as they currently typically use dedicated hardware and facilities, which are only used after a successful observation is obtained, further increasing the total carbon footprint of science.

*A. The Green Cloud Computing example*

From the perspective of Green Cloud Computation, there are several aspects that have been tackled in order to increase the efficiency and decrease OPEX of current infrastructures, such as location, infrastructure reuse, equipment selection (servers, racks, networking), and cooling parameters. With Moore's law, chip power densities double every 18-24 months and technology upgrade simply follows with a rapid increase of power consumption. For instance, Large Data centers networks are based on the so called commodity High Power Computing (HPC) system architectures hosting work-horse applications like search engines and popular social networks services and have influenced as well the design of HPC systems for scientific applications. Some of these HPC systems require heavy cooling installations, raising extraordinarily their total power profile and their cost of ownership (TCO). Power availability and consumption became the biggest concern rather than speed ("performance") leading big ICT providers to focus on power consumption trimming of their large data centers and HPC systems, facilities well known for their high energy demand quite comparable in many cases to a medium sized city [28]. These concerns led to the appearance of the Green 500 ranking listing the top most efficient, power aware supercomputers [33]. Also, big Internet providers operating large Green high-end computing systems such as Data Centers and Clouds allocate resources to applications hosting Internet to meet users' quality of service requirements and minimize power consumption [30].

Recently, the reuse of devices reaching their end-of-life has also been addressed as a way to reduce the ecological footprint of a given system. Green ICT led to reduction or elimination of underused servers, which results in lower energy usage, in line with policy recommendations [26]. Ultimately, if considering the operational stage of a datacenter, the most common metrics for evaluating the efficiency of a computational infrastructure are FLOPS per Watt (F/W) and Power User Efficiency (PUE), where PUE= (Total Facility Energy/ Information Equipment Energy). An ideal PUE value would be 1, with nearly all of the energy being used for computing whereas state-of-art is already PUE $\leq$ 1.2 for some greener large datacenters. A PUE value of 2.0 means that for every watt of IT equipment power, an additional watt is consumed to cool and distribute power to the IT equipment [30]. A quoted example, with the aid of a ultra-efficient evaporative cooling technology, Google Inc. has been able to reduce its energy consumption to 50% of that of the industry average with a very low PUE~1.1 [32]. Most of these metrics can also be applied to large Science Infrastructures. In addition some tasks may be off-loaded to public clouds having lower PUE values.

Location is a major aspect driving the development of a large computational cloud facility. Ideally, a datacenter should be placed next to a power source so that the price is minimum, and losses in the power grid are minimized (est. 17% is lost in the power grid [26]). If the project has ecological aspects, as it is common to observe, water dams, wind turbines and solar panels may be favored. Moreover, there must be interconnectivity to the global Internet through multiple providers. Climate and geography also play an important role with great impact in temperature control, and overall security of the infrastructure. Free cooling provided by rivers and oceans, and even wind, can be used if available.

However, for most large science facilities location is conditioned by the experiment, and not by the computational facilities, which results in far from optimal efficiency, higher capital expenditure (CAPEX) and higher OPEX. As an example, ALMA, with its correlator located in the middle of the Atacama Desert, at an altitude of 5km, far from power sources, and with a thin atmosphere, presents serious engineering challenges even for keeping basic operation, and just without addressing efficiency concerns.

Infrastructure reuse is another important aspect that is always considered, and at multiple levels. In the area of computing and Internet service provisioning, it is possible to increase the usage rate of computational resources (servers), by exploring virtualization and service-oriented technologies, mostly due to the intermittent resource consumption pattern shown by almost any application or service. By combining multiple, unrelated services in the same hardware resources, processing cycles can be multiplexed, ensuring that overcapacity is reduced to a minimum. Using this technique, servers are optimized and redesigned to become highly power efficient, with smart dynamic power management of consumption links profiles with real-time power-aware configuring ability of system devices, such as processors, disks, and communications. Some cloud providers, even have spot pricing for their resources, according to laws of demand and supply. In this aspect, private Cloud Computing technologies have emerged as a promising Green ICT solution, which can be exploited by Big Data Centers and Science Organizations [6]-[8]. In fact, Cloud computing is leading data center operators to trim energy costs and reduce carbon emissions, hence addressing also the management and power concerns of large scale science infrastructures. This led to a new concept, of Green Data Centers, combining clever strategies with Green Power availability, attractive by the enabled reduction of greenhouse gas emissions [19],[ 27],[28], [30]. Figure 1, depicts the evolution of greenhouse gas reduction projections in time related to servers-only. R&D is being done to reduce the undesirable effects e.g. in the so-called DOME project to be addressed later in this paper.

III. SCIENCE INFRASTRUCTURES : TOWARDS THE DATA DELUGE

*A. The Low Frequency Array (LOFAR)*

The LOFAR Telescope, officially launched in 2010, is a novel phased-array radio interferometer containing more than 10,000 small antennas distributed across the Netherlands, the UK, Germany, France and Sweden. The core of the array, consisting of some 40 stations, is located in the North-eastern part of the Netherlands; an additional eight stations are distributed over the participating countries [see e.g. 20]. The LOFAR telescope facilities are jointly operated by the International LOFAR Telescope (ILT) foundation. For training and other purposes in a living lab context four so-called remote stations in the Netherlands have been identified for potential enhancements to full solar-powered operation. Note that in the case of radio astronomy, the issue of radio frequency interference is a key lay-out and design issue.

That being so, the present LOFAR operation is designed around an intense real-time raw data stream of tens of Terabits/s reducing to roughly 150 Gbits/s after beamforming. This data stream is sent to the central processor (correlator), requiring partially dedicated fibre networks for long-range data transport. After correlation, typical imaging observations can easily produce visibility data at rates exceeding 35 TByte/h. After processing and analyzing, the data volumes of science products are reduced significantly, leading to an expected growth of 5 pByte per year for the Long-Term Archive (LTA).

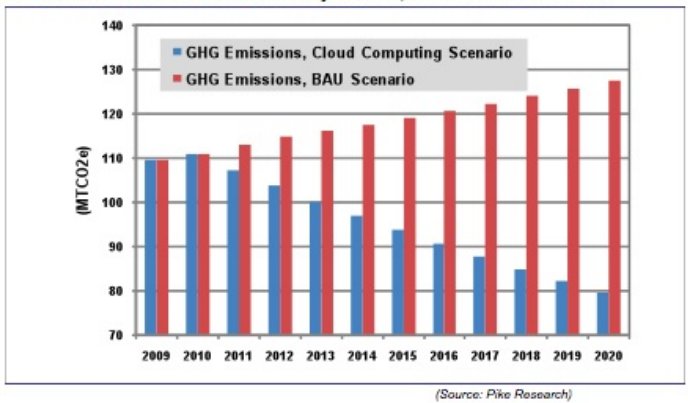

Fig1. Greenhouse Gas Emission reduction projections for Cloud Computing Scenarios [19] related to servers-only Reduction of undesirable effects are topic of active R&D.

*B. The Square Kilometre Array*

The Square Kilometre Array (SKA) is an international multipurpose next-generation radio interferometer, an Information and Communication Technology machine with thousands of antennas linked together to provide a collecting area of one square kilometer [3]. The SKA is the only global project in the European Strategy Forum of Research Infrastructures, with 10 Full members (Australia, Canada, China, Germany, Italy, New Zealand, South Africa, Sweden, The Netherlands and the United Kingdom) and an Associated member (India). It further involves more than 67 organizations in 20 countries, and counts with world-leading ICT industrial partners. The SKA will be built in the Southern Hemisphere in high solar irradiated zones (mainly in South Africa, with distant stations in the SKA African Partners - Botswana, Ghana, Kenya, Zambia, Madagascar, Mauritius, Mozambique, Namibia - and Australia/New Zealand). It can be best described a central core of ~200 km diameter, with 3 spiral arms of cables connecting nodes of antennas spreading over sparse territories in several countries up to 3000km distances, all in high solar irradiance latitudes. Creating a power grid covering several countries with a diameter of 3000km is impractical. It is also impractical to rely on fuel sources such as diesel, due to the remote location of most antennas. Solar Power supply is therefore an option to the power generation mix of the SKA antennas, and to contribute towards a zero carbon footprint during its lifetime, and reduced OPEX. Table 1 shows the estimated power needs of the SKA for the two installation sites. Current electronic technology projections point towards an expected target average power usage of approximately 100 MW [5]-[7], when combining all systems. Since the SKA will continuously scan the sky, it will not present strong power peaks and power-fluctuations, keeping a much smoother but demanding consumption profile. Energy generation at a continental scale for this facility, with different load profiles at different locations, means that modular power generators are needed, presenting an ideal scenario for development of innovative solutions with its own degree of customization and grid connectivity.

Figure 2 above depicts the core as part of a network of receptors of about 3000 reflector antennas and several million so-called aperture array antennas along three arms of several hundreds kilometers. SKA is planned in two construction Phases, with deployment of different sensor technologies. In particular, performance of digital sensors [1],[6] may be driven by the electronics power consumption, as power consumption may likely cap sensor performance with a direct impact on system sensitivity. To cut consumption, innovative forms of passive cooling, plus room temperature operation, will be considered for the work of the Low Noise Amplifiers of the antenna sensors. Therefore, to extract the maximum scientific potential and maintain costs at appropriate levels it is essential to couple the power cycles with electronic power requirements (power and cooling), in a hot, dry site, with temperatures closer to 50ºC in open field for viable operation.

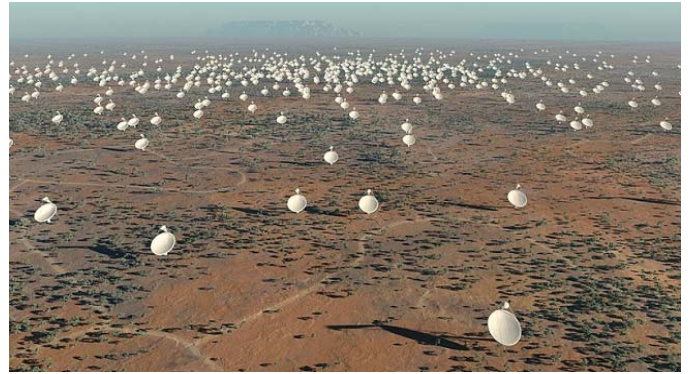

Fig2. An artist vision of the SKA Core site, with some of the projected 3000 15-meter parabolic dishes. From http://www.skatelescope.org. Together with several million so-called aperture array antennas alike employed in (e.g. LOFAR), the core with receptors along networks of three arms stretching out to hundreds of kilometres, constitute a data intense and high performance compute exa-scale level ICT based infrastructure.

This should be done without degrading receiver system temperature to achieve expected sensitivities, detailed in the SKA Key Science Projects.

The SKA will have different operational modes, each requiring a different computational load and power, and a different distribution of this power over the subsystems. Therefore, not only the power consumption of the SKA as a whole will vary over time, also the spatial distribution of the power consumption over the SKA systems will vary. Other aspects that may influence the distribution of power consumption are error-induced re-routing of data-streams, and future upgrades and system extensions. On the supply-side there will be additional constraints such as the interrupted availability of solar irradiance and limited availability of energy storage capacity. This clearly implies that a Smart Grid approach, both at system level and subsystem level, is needed in order to reduce the carbon footprint of the telescope.

Overall, the main characteristic concerning the SKA power system can summarized as:

- Many Antennas nodes are far away from civilization centers and power grid in climates with high thermal amplitudes.

- Exquisite control of Radio Frequency Interference and EMI from Power systems is needed, since RFI would impair the radio telescope sensitivity.
- Different Power requirements over large distances: the SKA Core will require approximately 50MW; the High power Computing <50M€; around 80 stations of about 100kW over the spiral arms.
- Continuous operation (meaning 24/7 availability) for sky surveying points out that some serious storage capabilities are required, and power supply for night operations must be carefully considered.
- Power stability: control of current peaks, for operation, cooling, computing and telescope management and monitoring.
- Scalability: the power infrastructure should scale from SKA Phase 1 to the later, more extended, and more power demanding Phase2.
- Data output: ~1-10 Exabyte per day for the correlator to process; 1Petabyte of data per day for further offline reduction and analysis.

*C. The Dome project*

The next generation large-scale scientific instruments, such as the SKA, require signal and data processing capacities exceeding current state-of-the art technologies. The DOME project [21] is a 5-year collaboration between ASTRON and IBM, aimed at developing emerging technologies for large-scale and efficient (green) exascale computing, data transport, storage, and streaming processing. Based on experience gained with a retrospective analysis of LOFAR, the DOME team analyzed the compute and power requirements of the telescope concepts for the first phase of the SKA [23]. These initial estimates indicate that the power requirements are challenging (up to order ten peta operations per second (OPS) in the station processing and correlation), but especially the post correlation processing (order 100 peta OPS to exa OPS) is dominating the power consumption [22],[23]. The study also poses mitigation strategies, such as developing more efficient algorithms, fine-tuning the calibration and imaging processing parameters, and phased-implementation of novel accelerator technologies.

IV. CONCLUSION

The reduction of radioastronomy large infrastructures carbon footprint requires the development of innovative strategies combining power efficiency gains in electronics and intelligent data center management along the trend of Green ICT: towards Greener Processing and Storage coupled with availability of a Greener power mix. Solar Energy is being investigated as one of the power supply options to mega science projects in remote, highly irradiated locations. This will require SmartGrid management of the available power mix at system and subsystem level. Among several power option investigations, already some demo projects like the BIOSTIRLING-4SKA are studying the viability of new provision technologies like Stirling Dish-engine-based pilot plant for a 24/7 operation of radio astronomical prototype technologies, in Portugal. These developments will set a new paradigm in sustainability in large scale radio astronomy infrastructures

V. ACKNOWLEDGMENTS

LOFAR is funded though the BSIK program, EFRO, SNN, and N.W.O. DOME is a Public Partnership between IBM and ASTRON, supported by the Dutch Ministry of Economic Affairs and the Province of Drenthe. VIA-SKA is funded by Ministerio de Economia y Competitividad under Grant AIC-A-2011-0658. LVM and JSV acknowledge AYA2011-30491-C02-01, grant co-financed by MICINN and FEDER funds, and the Junta de Andalucía (Spain) grant P08-FQM-4205. DB acknowledges GEM grant support from IT and sponsorship by TICE and ENGAGE SKA Consortium. JSV also acknowledges support for his JAE-Doc contract, cofounded by CSIC and the European Social Fund. AvA, DB, JSV, LVM acknowledge support of the BIOSTIRLING 4 SKA project, funded under FP7 ENERGY.2012.2.5-1, project number 309028.